\documentclass[10pt,conference,hidelinks]{IEEEtran}

\usepackage{cite}
\usepackage{amsmath,amssymb,amsfonts}
\usepackage{siunitx}
\usepackage{graphicx}
\usepackage{textcomp}
\usepackage{xcolor}
\def\BibTeX{{\rm B\kern-.05em{\sc i\kern-.025em b}\kern-.08em
    T\kern-.1667em\lower.7ex\hbox{E}\kern-.125emX}}
\usepackage{glossaries}
\usepackage{orcidlink}
\usepackage[position=b]{subcaption}
\usepackage{cleveref}
\usepackage{tikz}
\usetikzlibrary{tikzmark,calc,positioning}
\usepackage{algorithm}
\usepackage{algpseudocode}
\usepackage{listings}
\usepackage{makecell}
\usepackage{booktabs}
\usepackage{multirow}
\usepackage{siunitx}
\usepackage{threeparttable}

\newacronym[plural=WANs, firstplural={Wide Area Networks (WANs)}]{wan}{WAN}{Wide Area Network}
\newacronym[plural=WSNs, firstplural={Wireless Sensor Networks (WSNs)}]{wsn}{WSN}{Wireless Sensor Network}
\newacronym{simd}{SIMD}{Single Instruction Multiple Data}
\newacronym{os}{OS}{Operating System}
\newacronym{ble}{BLE}{Bluetooth Low-Energy}
\newacronym{wifi}{Wi-FI}{Wireless Fidelity}
\newacronym[plural=DVS, firstplural={Dynamic Vision Sensors (DVS)}]{dvs}{DVS}{Dynamic Vision Sensor}
\newacronym{ptz}{PTZ}{Pan-Tilt Unit}
\newacronym{ge}{GE}{gate equivalent}

\newacronym[plural=FLLs,firstplural=Frequency Locked Loops (FLLs)]{fll}{FLL}{Frequency Locked Loop}
\newacronym{dram}{DRAM}{Dynamic Random Access Memory}
\newacronym{fpu}{FPU}{floating-point unit}
\newacronym{fpss}{FPSS}{Floating Point Subsystem}
\newacronym{frep}{FREP}{Floating Point Repetition}
\newacronym{dma}{DMA}{direct memory access}
\newacronym{ssr}{SSR}{stream semantic register}
\newacronym{issr}{ISSR}{Indirection Stream Semantic Register}
\newacronym[plural=LUTs, firstplural={Lookup Tables (LUTs)}]{lut}{LUT}{Lookup Table}
\newacronym[plural=FPGAs, firstplural={Field Programmable Gate Arrays (FPGAs)}]{fpga}{FPGA}{Field Programmable Gate Array}
\newacronym{dsp}{DSP}{digital signal processor}
\newacronym{mcu}{MCU}{Microcontroller Unit}
\newacronym{spi}{SPI}{Serial Peripheral Interface}
\newacronym{cpi}{CPI}{Camera Parallel Interface}
\newacronym{rf}{RF}{register file}
\newacronym{fifo}{FIFO}{First-In First-Out Queue}
\newacronym{uart}{UART}{Universal Asynchronous Receiver-Transmitter}
\newacronym{raw}{RAW}{read after write}
\newacronym[plural=ISAs, firstplural={instruction set architectures (ISAs)}]{isa}{ISA}{instruction set architecture}
\newacronym{xbar}{XBAR}{crossbar}
\newacronym[firstplural=Scratch-Pad Memories (SPMs)]{spm}{SPM}{Scratch-Pad Memory}
\newacronym{ppa}{PPA}{Power Performance Area}
\newacronym{ipi}{IPI}{Inter-Processor Interrupt}
\newacronym[firstplural=Software-Generated Interrupts (SGIs)]{sgi}{SGI}{Software-Generated Interrupt}
\newacronym{pe}{PE}{processing element}
\newacronym{tcdm}{TCDM}{Tightly-Coupled Data Memory}
\newacronym{lsu}{LSU}{Load-Store Unit}
\newacronym{icache}{I\$}{Instruction Cache}
\newacronym{dcache}{D\$}{Data Cache}
\newacronym{wfi}{WFI}{Wait For Interrupt}
\newacronym{gpc}{GPC}{GPU Processing Cluster}
\newacronym{cpu}{CPU}{Central Processing Unit}
\newacronym{gpu}{GPU}{Graphics Processing Unit}
\newacronym{llc}{LLC}{Last-Level Cache}
\newacronym{sm}{SM}{Streaming Multiprocessor}
\newacronym[firstplural=Networks on Chip (NoCs)]{noc}{NoC}{Network on Chip}
\newacronym{dsa}{DSA}{Domain Specific Accelerator}
\newacronym{rb}{RB}{ring buffer}
\newacronym{msb}{MSB}{most significant bit}

\newacronym{dfg}{DFG}{Data Flow Graph}
\newacronym{lcg}{LCG}{Linear Congruential Generator}
\newacronym{prn}{PRN}{Pseudo-Random Number}
\newacronym{matmul}{matmul}{matrix multiplication}

\newacronym{ste}{STE}{Straight-Through-Estimator}

\newacronym[plural=PTUs, firstplural={Pan-Tilt Units}]{ptu}{PTU}{Pan-Tilt Unit}
\newacronym{mdf}{MDF}{Medium-density fibreboard}
\newacronym{cvat}{CVAT}{Computer Vision Annotation Tool}
\newacronym{coco}{COCO}{Common Objects in Context}
\newacronym{soa}{SoA}{state-of-the-art}
\newacronym{sf}{SF}{Sensor Fusion}

\newacronym{dl}{DL}{Deep Learning}
\newacronym{bn}{BN}{Batch Normalization}
\newacronym{FGSM}{FBK}{Fast Gradient Sign Method}
\newacronym{lr}{LR}{Learning Rate}
\newacronym{sgd}{SGD}{Stochastic Gradient Descent}
\newacronym{gd}{GD}{Gradient Descent}
\newacronym{llm}{LLM}{large language model}

\newacronym{sta}{STA}{Static Timing Analysis}

\newacronym[plural=GPIOs, firstplural={General Purpose Inupt Outputs (GPIOs)}]{gpio}{GPIO}{General Purpose Input Output}
\newacronym[plural=LDOs, firstplural={Low Dropout Regulators (LDOs)}]{ldo}{LDO}{Low Dropout Regulator}

\newacronym{inq}{INQ}{Incremental Network Quantization}

\newacronym{CV}{CV}{Computer Vision}
\newacronym{EoT}{EoT}{Expectation over Transformation}
\newacronym{RPN}{RPN}{Region Proposal Network}
\newacronym{TV}{TV}{Total Variation}
\newacronym{NPS}{NPS}{Non-Printability Score}
\newacronym{STN}{STN}{Spatial Transformer Network}
\newacronym{MTCNN}{MTCNN}{Multi-Task Convolutional Neural Network}
\newacronym{YOLO}{YOLO}{You Only Look Once}
\newacronym{SSD}{SSD}{Single Shot Detector}
\newacronym{NMS}{NMS}{Non-Maximum Suppression}
\newacronym{ic}{IC}{Integrated Circuit}
\newacronym{tcxo}{TCXO}{Temperature Controlled Crystal Oscillator}
\newacronym{jtag}{JTAG}{Joint Test Action Group industry standard}
\newacronym{swd}{SWD}{Serial Wire Debug}
\newacronym{sdio}{SDIO}{Serial Data Input Output}

\newacronym[plural=PCBs, firstplural={Printed Circuit Boards (PCB)}]{pcb}{PCB}{Printed Circuit Board}
\newacronym[plural=ASICs, firstplural={Application Specific Integrated Circuits}]{asic}{ASIC}{Application Specific Integrated Circuit}

\newacronym[plural=BNNs, firstplural={Binary Neural Networks (BNNs)}]{bnn}{BNN}{Binary Neural Network}
\newacronym[plural=NNs, firstplural={Neural Networks}]{nn}{NN}{Neural Network (NNs)}
\newacronym[plural=SCMs, firstplural={Standard Cell Memories (SCMs)}]{scm}{SCM}{Standard Cell Memory}
\newacronym{ann}{ANN}{Artificial Neural Networks}
\newacronym{ml}{ML}{machine learning}
\newacronym{ai}{AI}{Artificial Intelligence}
\newacronym{iot}{IoT}{Internet of Things}
\newacronym{fft}{FFT}{Fast Fourier Transform}
\newacronym[plural=OCUs, firstplural={Output Channel Compute Units (OCUs)}]{ocu}{OCU}{Output Channel Compute Unit}
\newacronym{alu}{ALU}{Arithmetic Logic Unit}
\newacronym{mac}{MAC}{Multiply-Accumulate}
\newacronym[firstplural={systems-on-chip (SoCs)}]{soc}{SoC}{system-on-chip}
\newacronym[firstplural={multi-processor systems-on-chip (MPSoCs)}]{mpsoc}{MPSoC}{multi-processor system-on-chip}

\newacronym{PGD}{PGD}{Projected Gradient Descend}
\newacronym{CW}{CW}{Carlini-Wagner}
\newacronym{OD}{OD}{Object Detection}

\newacronym{rrf}{RRF}{RADAR Repetition Frequency}
\newacronym{nlp}{NLP}{Natural Language Processing}
\newacronym{qam}{QAM}{Quadrature Amplitude Modulation}
\newacronym{rri}{RRI}{RADAR Repetition Interval}
\newacronym{radar}{RADAR}{Radio Detection and Ranging}
\newacronym{loocv}{LOOCV}{Leave-one-out cross validation}

\newacronym{bsp}{BSP}{Board Support Package}
\newacronym{ttn}{TTN}{The Things Network}
\newacronym{wip}{WIP}{Work in Progress}
\newacronym{json}{JSON}{JavaScript Object Notation}
\newacronym{qat}{QAT}{Quantization-Aware Training}

\newacronym{cls}{CLS}{Classification Error}
\newacronym{loc}{LOC}{Localization Error}
\newacronym{bkgd}{BKGD}{Background Error}
\newacronym{roc}{ROC}{Receiver Operating Characteristic}
\newacronym{frr}{FRR}{False Rejection Rate}
\newacronym{eer}{EER}{Equal Error Rate}
\newacronym{snr}{SNR}{Signal-to-Noise Ratio}
\newacronym{flop}{FLOP}{Floating-Point Operation}
\newacronym{fp}{FP}{Floating-Point}
\newacronym{fps}{FPS}{Frames Per Second}
\newacronym{oi}{OI}{Operational Intensity}
\newacronym{ipc}{IPC}{Instructions per Cycle}

\newacronym{gsc}{GSC}{Google Speech Commands}
\newacronym{mswc}{MSWC}{Multilingual Spoken Words Corpus}
\newacronym{demand}{DEMAND}{Diverse Environments Multichannel Acoustic Noise Database}

\newacronym[plural=SNNs, firstplural={Spiking Neural Networks (SNNs)}]{snn}{SNN}{Spiking Neural Network}
\newacronym[plural=DNNs, firstplural={deep neural networks (DNNs)}]{dnn}{DNN}{deep neural network}
\newacronym[plural=TCNs,firstplural=Temporal Convolutional Networks]{tcn}{TCN}{Temporal Convolutional Network}
\newacronym[plural=CNNs,firstplural=Convolutional Neural Networks (CNNs)]{cnn}{CNN}{Convolutional Neural Network}
\newacronym[plural=TNNs,firstplural=Ternarized Neural Networks]{tnn}{TNN}{Ternarized Neural Network}
\newacronym{ds-cnn}{DS-CNN}{Depthwise Separable Convolutional Neural Network}
\newacronym{rnn}{RNN}{Recurrent Neural Network}
\newacronym{gcn}{GCN}{Graph Convolutional Network}
\newacronym{mhsa}{MHSA}{Multi-Head Self Attention}
\newacronym{crnn}{CRNN}{Convolutional Recurrent Neural Network}
\newacronym{clca}{CLCA}{Convolutional Linear Cross-Attention}

\newacronym{bf}{BF}{Beamforming}
\newacronym{anc}{ANC}{Active Noise Cancellation}
\newacronym{agc}{AGC}{Automatic Gain Control}
\newacronym{se}{SE}{Speech Enhancement}
\newacronym{mct}{MCT}{Multi-Condition Training}
\newacronym{mcta}{MCTA}{Multi-Condition Training \& Adaptation}
\newacronym{pcen}{PCEN}{Per-Channel Energy Normalization}
\newacronym{mfcc}{MFCC}{Mel-Frequency Cepstral Coefficient}
\newacronym{asr}{ASR}{Automated Speech Recognition}
\newacronym{kws}{KWS}{Keyword Spotting}
\newacronym{odl}{ODL}{On-Device Learning}

\newacronym{nl-kws}{NL-KWS}{Noiseless Keyword Spotting}
\newacronym{na-kws}{NA-KWS}{Noise-Aware Keyword Spotting}
\newacronym{odda}{ODDA}{On-Device Domain Adaptation}
\newacronym{hpm}{HPM}{High-Performance Mode}
\newacronym{lpm}{LPM}{Low-Power Mode}

\newcommand{\ResultBaseThirtyTwoFcMinUtilization}{78.5}
\newcommand{\ResultBaseThirtyTwoFcMaxUtilization}{94.0}
\newcommand{\ResultBaseThirtyTwoFcMedianUtilization}{88.2}

\newcommand{\ResultZonlThirtyTwoFcMedianUtilization}{93.4}
\newcommand{\ResultZonlSixtyFourFcMinUtilization}{88.9}

\newcommand{\ResultZonlSixtyFourFcMedianUtilization}{98.1}
\newcommand{\ResultZonlSixtyFourFcMedianUtilizationOverZonlThirtyTwoFc}{5}
\newcommand{\ResultZonlSixtyFourFcBottomWhiskerUtilization}{96.2}
\newcommand{\ResultZonlSixtyFourFcTopWhiskerUtilization}{99.4}

\newcommand{\ResultZonlSixtyFourDobuMedianUtilizationOverZonlThirtyTwoFc}{5}

\newcommand{\ResultZonlFourtyEightDobuMedianUtilizationOverBaseThirtyTwoFc}{11}
\newcommand{\ResultZonlFourtyEightDobuBottomWhiskerUtilization}{96.1}
\newcommand{\ResultZonlFourtyEightDobuTopWhiskerUtilization}{99.4}
\newcommand{\ResultZonlThirtyTwoFcMedianPowerOverBaseThirtyTwoFc}{4}
\newcommand{\ResultZonlSixtyFourDobuMedianPowerOverZonlThirtyTwoFc}{6}
\newcommand{\ResultZonlSixtyFourFcMedianEnergyOverZonlThirtyTwoFc}{12}

\newcommand{\ResultZonlFourtyEightDobuMedianEnergyOverBaseThirtyTwoFc}{8}

\newcommand{\ResultBaseThirtyTwoCellArea}{3.75}
\newcommand{\ResultBaseThirtyTwoMacroArea}{1.51}
\newcommand{\ResultBaseThirtyTwoTotalArea}{5.26}
\newcommand{\ResultBaseThirtyTwoWireLength}{26.6}
\newcommand{\ResultZonlThirtyTwoFcCellArea}{3.90}
\newcommand{\ResultZonlThirtyTwoFcMacroArea}{1.51}
\newcommand{\ResultZonlThirtyTwoFcTotalArea}{5.41}
\newcommand{\ResultZonlThirtyTwoFcWireLength}{27.4}
\newcommand{\ResultZonlSixtyFourFcCellArea}{4.67}
\newcommand{\ResultZonlSixtyFourFcMacroArea}{1.81}
\newcommand{\ResultZonlSixtyFourFcTotalArea}{6.48}
\newcommand{\ResultZonlSixtyFourFcWireLength}{34.8}
\newcommand{\ResultZonlSixtyFourDobuCellArea}{4.09}
\newcommand{\ResultZonlSixtyFourDobuMacroArea}{1.81}
\newcommand{\ResultZonlSixtyFourDobuTotalArea}{5.90}
\newcommand{\ResultZonlSixtyFourDobuWireLength}{29.3}
\newcommand{\ResultZonlFourtyEightDobuCellArea}{3.92}
\newcommand{\ResultZonlFourtyEightDobuMacroArea}{1.39}
\newcommand{\ResultZonlFourtyEightDobuTotalArea}{5.32}
\newcommand{\ResultZonlFourtyEightDobuWireLength}{26.6}
\newcommand{\ResultZonlSixtyFourFcCellAreaIncreaseOverZonlThirtyTwoFc}{14.3}
\newcommand{\ResultZonlSixtyFourFcWireLengthIncreaseOverZonlThirtyTwoFc}{27.1}
\newcommand{\ResultZonlSixtyFourFcMacroAreaIncreaseOverZonlThirtyTwoFc}{5.4}
\newcommand{\ResultZonlSixtyFourDobuCellAreaIncreaseOverZonlThirtyTwoFc}{3.6}
\newcommand{\ResultZonlSixtyFourDobuWireLengthIncreaseOverZonlThirtyTwoFc}{7.1}
\newcommand{\ResultZonlFourtyEightDobuCellAreaIncreaseOverZonlThirtyTwoFc}{0.5}
\newcommand{\ResultZonlFourtyEightDobuWireLengthIncreaseOverZonlThirtyTwoFc}{2.9}
\newcommand{\ResultZonlFourtyEightDobuMacroAreaIncreaseOverZonlThirtyTwoFc}{8.1}
\newcommand{\ResultZonlFourtyEightDobuTotalAreaIncreaseOverZonlThirtyTwoFc}{1.8}
\newcommand{\ResultZonlThirtyTwoFcTotalAreaOverBaseThirtyTwoFc}{3}
\newcommand{\ResultZonlSixtyFourFcTotalAreaOverBaseThirtyTwoFc}{23}
\newcommand{\ResultZonlSixtyFourDobuTotalAreaOverBaseThirtyTwoFc}{12}
\newcommand{\ResultZonlFourtyEightDobuTotalAreaOverBaseThirtyTwoFc}{1}
\newcommand{\ResultZonlThirtyTwoFcWireLenOverBaseThirtyTwoFc}{3}
\newcommand{\ResultZonlSixtyFourFcWireLenOverBaseThirtyTwoFc}{31}
\newcommand{\ResultZonlSixtyFourDobuWireLenOverBaseThirtyTwoFc}{10}
\newcommand{\ResultZonlFourtyEightDobuWireLenOverBaseThirtyTwoFc}{-0.2}
\newcommand{\ResultZonlFourtyEightDobuFpuArea}{1.51}
\newcommand{\ResultZonlFourtyEightDobuTcdmArea}{1.26}
\newcommand{\ResultZonlFourtyEightDobuInterconnectArea}{0.91}
\newcommand{\ResultZonlFourtyEightDobuControlArea}{2.56}

\newcommand{\ResultBaseThirtyTwoFcFpuArea}{1.48}
\newcommand{\ResultBaseThirtyTwoFcTcdmArea}{1.38}
\newcommand{\ResultBaseThirtyTwoFcInterconnectArea}{0.92}
\newcommand{\ResultBaseThirtyTwoFcControlArea}{2.41}
\newcommand{\ResultBaseThirtyTwoFcTotalArea}{5.26}
\newcommand{\ResultOpenGemmTcdmArea}{2.44}
\newcommand{\ResultOpenGemmComputeArea}{1.43}
\newcommand{\ResultOpenGemmControlArea}{0.86}
\newcommand{\ResultOpenGemmTotalArea}{3.85}
\newcommand{\ResultOpenGemmFpuPower}{106.3}
\newcommand{\ResultOpenGemmTcdmPower}{90.2}
\newcommand{\ResultOpenGemmControlPower}{93.0}
\newcommand{\ResultOpenGemmTotalPower}{289.5}
\newcommand{\ResultOpenGemmUtilization}{95}
\newcommand{\ResultOpenGemmPerformance}{7.60}
\newcommand{\ResultOpenGemmAreaEfficiency}{16.3}
\newcommand{\ResultOpenGemmEnergyEfficiency}{26.3}
\newcommand{\ResultBaseThirtyTwoFcUtilization}{95.3}
\newcommand{\ResultBaseThirtyTwoFcPerformance}{7.63}
\newcommand{\ResultBaseThirtyTwoFcAreaEfficiency}{12.0}
\newcommand{\ResultBaseThirtyTwoFcEnergyEfficiency}{22.4}
\newcommand{\ResultZonlFourtyEightDobuUtilization}{99.0}
\newcommand{\ResultZonlFourtyEightDobuPerformance}{7.92}
\newcommand{\ResultZonlFourtyEightDobuAreaEfficiency}{12.4}
\newcommand{\ResultZonlFourtyEightDobuEnergyEfficiency}{23.2}

\newcommand{\ResultZonlFourtyEightDobuEnergyEfficiencyOverOpenGemm}{12}
\newcommand{\ResultBaseThirtyTwoFcFpuPower}{106.7}
\newcommand{\ResultZonlFourtyEightDobuFpuPower}{115.0}
\newcommand{\ResultBaseThirtyTwoFcTcdmPower}{47.5}
\newcommand{\ResultZonlFourtyEightDobuTcdmPower}{36.9}
\newcommand{\ResultBaseThirtyTwoFcControlPower}{186.3}
\newcommand{\ResultZonlFourtyEightDobuControlPower}{189.2}
\newcommand{\ResultBaseThirtyTwoFcTotalPower}{340.4}
\newcommand{\ResultZonlFourtyEightDobuTotalPower}{341.1}

\makeatletter
\newcommand{\insertnewlines}[1]{%
  \noindent\mbox{}%
  \@tempcnta=#1\relax
  \loop\ifnum\@tempcnta>0
    \\
    \advance\@tempcnta\m@ne
  \repeat
}
\makeatother

\DeclareSIUnit{\reads}{reads}
\DeclareSIUnit{\write}{write}
\DeclareSIUnit{\cores}{cores}
\DeclareSIUnit{\hyperbanks}{hyperbanks}
\DeclareSIUnit{\banks}{banks}
\DeclareSIUnit{\getsmc}{GE_{TSMC16}}
\DeclareSIUnit{\gegf}{GE_{GF12}}

\begin{document}

\title{Towards Zero-Stall Matrix Multiplication on\\Energy-Efficient RISC-V Clusters for\\Machine Learning Acceleration}

\ifdefined\blindreview
\else
\author{
    \IEEEauthorblockN{
        Luca Colagrande\orcidlink{0000-0002-7986-1975}\IEEEauthorrefmark{1}, 
        Lorenzo Leone\orcidlink{0009-0000-3976-847X}\IEEEauthorrefmark{1}, 
        Maximilian Coco\orcidlink{0009-0008-0820-5576}\IEEEauthorrefmark{2}, 
        Andrei Deaconeasa\orcidlink{0009-0005-1410-7540}\IEEEauthorrefmark{2} 
        and Luca Benini\orcidlink{0000-0001-8068-3806}\IEEEauthorrefmark{1}
    }
    \IEEEauthorblockA{
        \IEEEauthorrefmark{1}Integrated Systems Laboratory (IIS), ETH Zurich, Zurich, Switzerland\\
        \IEEEauthorrefmark{2}D-ITET, ETH Zurich, Zurich, Switzerland\\
        \{colluca,lleone,lbenini\}@iis.ee.ethz.ch  \{mcoco,adeaconeasa\}@student.ethz.ch\\
    }
}
\fi

\maketitle

\begin{abstract}
    The growing computational demands of machine learning (ML) workloads have driven the design of ML accelerators aiming at an optimal tradeoff between efficiency and flexibility. A widely explored architecture for flexible ML accelerators is based on clusters of lightweight instruction processors sharing multi-banked L1 memory, augmented with specialized instruction extensions for key ML-related computations, such as matrix multiplication (matmul). However, instruction extensions should be coupled with microarchitectural optimizations that remove inefficiencies due to control flow (loop handling) and memory access, without drastically increasing processor complexity. Moving from a state-of-the-art (SoA) ML accelerator cluster based on RISC-V processors, we propose a low-overhead optimized microarchitecture that eliminates these inefficiencies almost entirely while retaining programmability. We introduce ``zero-overhead loop nests'' to remove control overheads, and a ``zero-conflict memory subsystem'', leveraging a novel double-buffering-aware interconnect, to eliminate bank conflicts in L1 memory. With these enhancements, we attain near-ideal utilizations between 96.1\% and 99.4\%,
    achieving 11\% performance and 8\% energy efficiency improvements over the baseline SoA RISC-V cluster. We demonstrate comparable utilizations and performance to a specialized SoA accelerator, with only 12\% difference in energy efficiency, while providing a fully-programmable general-purpose solution supporting a significantly wider range of workloads.
\end{abstract}
\begin{IEEEkeywords}
RISC-V, matrix multiplication, zero-overhead loops, bank conflicts, energy efficiency, general purpose
\end{IEEEkeywords}

\section{Introduction}

The computational demands of \gls{ml} workloads continue to grow at an unprecedented pace, far outstripping the capabilities of traditional general-purpose processors.
This surge has driven architects toward specialized tensor accelerators that can deliver superior performance and energy efficiency on dense linear algebra workloads \cite{jouppi2017}.

This approach however comes at the cost of flexibility, as \glspl{pe} allocated to fixed-function units can be exploited only on the targeted workloads.
This can result in lower overall application performance and area-efficiency compared to accelerators based on highly programmable \glspl{pe} \cite{nowatzki2016}.
On the other hand, programmable accelerators (e.g. GPUs) can employ their \glspl{pe} on a wide range of tasks, but this programmability comes at the cost of area and energy required by the control logic coupled to the \glspl{pe}.

A streamlined, low-overhead \gls{pe} microarchitecture is essential to develop general-purpose solutions that can approach the energy efficiency of specialized accelerators on \gls{ml} workloads.
``How close general-purpose accelerators can get to the energy efficiency of specialized accelerators?'' and ``how to reach this target?'' are open research questions.

RISC-V plays a pivotal role in answering these questions, as its open and extensible ISA enables open-source \gls{pe} design, optimization and benchmarking.
In a previous work, Zaruba et al. \cite{zaruba2021} developed an optimized general-purpose RISC-V multicore cluster for energy-efficient processing of floating-point intensive workloads.
They introduce Snitch, a tiny single-issue in-order RV32I control core coupled to a high-performance 64-bit SIMD-capable \gls{fpu}, providing a low-cost programming interface to the energy-hungry \gls{fpu}.

Key to achieve high energy efficiency in Snitch are two flexible ISA extensions: \textit{stream registers} \cite{domingos2021} and \textit{zero-overhead loops} \cite{kavvadias2008}, respectively implemented by Snitch's \glspl{ssr}  and FREP extensions.
Despite its simple and small in-order architecture, thanks to these extensions Snitch achieves a notable 85\% \gls{fpu} utilization on a 32$\times$32 double-precision (DP) \gls{matmul} kernel, in an 8-core cluster configuration evaluated by the authors.

The OpenGeMM platform \cite{yi2025} is an example of how to couple fixed-function units with processors to increase efficiency.
It couples a Snitch core with a \gls{matmul} accelerator to boost the energy efficiency up to a peak 4.68 INT8 Top/s/W.
Featuring custom control and data orchestration logic, it can achieve utilizations up to 99.34\% across various \gls{dnn} workloads.
While differences in arithmetic precision, technology and clock frequency prevent a direct comparison between the energy efficiency of the two platforms, \gls{pe} utilization figures can be used as an indicator of relative efficiency: there clearly is a utilization gap (99.34\% vs. 85\%) between  OpenGeMM and ``baseline'' Snitch. We aim at closing the gap, to maximize efficiency at a much lower area cost than what is implied by adding a dedicated \gls{matmul} accelerator.

In this paper, we propose an optimized general-purpose architecture building on Zaruba's cluster \cite{zaruba2021}, to close the gap with OpenGeMM and similar \gls{ml} accelerators.
Specifically:
\begin{itemize}
    \item We evaluate \gls{matmul} bottlenecks on the energy-efficient \gls{soa} general-purpose Snitch cluster \cite{zaruba2021}.
    We leverage its open source nature to pinpoint utilization losses in cycle-accurate RTL simulation, enabling direct correlation to microarchitectural details.
    \item We propose an optimized cluster architecture which eliminates \gls{matmul} inefficiencies with a minimal area overhead, while retaining programmability.
    This is achieved by means of two general-purpose extensions to the Snitch cluster, namely \textit{zero-overhead loop nests}, eliminating outer loop overheads, and a novel \textit{zero-conflict memory subsystem} optimized for double-buffered applications, eliminating bank conflicts.
    \item We evaluate our work on a key kernel for \gls{ml} workloads, matrix multiplication, achieving near-ideal utilizations between \ResultZonlFourtyEightDobuBottomWhiskerUtilization\% and \ResultZonlFourtyEightDobuTopWhiskerUtilization\% across various problem sizes, improving the median performance and energy efficiency of the heavily-optimized Snitch cluster by \ResultZonlFourtyEightDobuMedianUtilizationOverBaseThirtyTwoFc\% and \ResultZonlFourtyEightDobuMedianEnergyOverBaseThirtyTwoFc\%, respectively.
    We achieve comparable utilization and performance to OpenGeMM \cite{yi2025}, a specialized \gls{soa} accelerator, with only a limited \ResultZonlFourtyEightDobuEnergyEfficiencyOverOpenGemm\% difference in energy efficiency, while enabling our \glspl{pe} to be employed on a significantly wider range of workloads.
\end{itemize}

\noindent
Our implementation is fully open-source and performance experiments are reproducible using free software.
\ifdefined\blindreview
    \footnote{https://hidden-for-double-blind-review.com}
\else
    \footnote{\url{https://github.com/colluca/snitch\_cluster/tree/islped}}
\fi

\section{Background}
\label{sec:background}

\begin{figure}[t]
    \centering
    \begin{subfigure}{0.55\columnwidth}
        \centering
        \includegraphics[width=\textwidth]{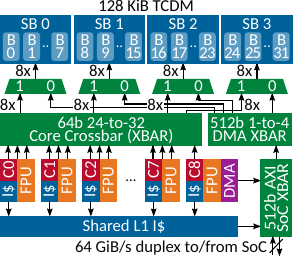}
        \caption{}
        \label{fig:snitch_cluster}
    \end{subfigure}
    \hfill
    \begin{subfigure}{0.435\columnwidth}
        \centering
        \includegraphics[width=\textwidth]{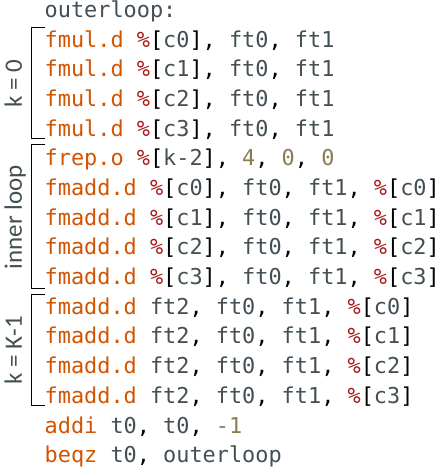}
        \caption{}
        \label{fig:gemm}
    \end{subfigure}
    \caption{(a) Architecture of the baseline Snitch cluster. (b) Optimized GEMM kernel for the baseline Snitch cluster.}
\end{figure}

\Cref{fig:snitch_cluster} shows the architecture of the baseline Snitch cluster, in its silicon-proven configuration from \cite{scheffler2025}.
It features 8 Snitch cores dedicated to compute tasks, and an additional core coupled with a \gls{dma} engine for data movement (DM) tasks.
All cores share a 128\,KiB software-managed L1 data cache, known as tightly-coupled data memory (TCDM).
The TCDM is partitioned into 32 banks, providing a peak bandwidth of 32 64-bit words per cycle, and adopts an interleaved address mapping across banks.
Every compute core features 3 64-bit ports to the TCDM, with access to every bank through a fully-connected interconnect.
The \gls{dma} features a 512-bit port for high-bandwidth burst-based transfers.
Through a dedicated interconnect branch, it can access any arbitrary superbank, i.e. a group of 8 contiguous banks, every cycle.
A mux at each superbank arbitrates the requests from the core and \gls{dma} interconnect branches.

Due to its significance for \gls{ml} workloads, we demonstrate our work on \gls{matmul}, computing $\mathbf{C} = \mathbf{A} \times \mathbf{B}$, given matrices $\mathbf{A}$, $\mathbf{B}$ and $\mathbf{C}$ of size M$\times$K, K$\times$N and M$\times$N, respectively.
\Cref{fig:gemm} presents a highly-optimized matmul code for Snitch, leveraging \glspl{ssr} and FREP.
The outer loop iterates over the elements of $\mathbf{C}$, and is partially unrolled to hide \gls{raw} stalls, resulting in 4 dot product computations (in registers \lstinline{c0}-\lstinline{c3}) every iteration\footnote{For brevity constraints, we adopt an unroll factor of 4 in our presentation. Actual implementations employ a factor of 8.}; specifically, between one row of $\mathbf{A}$ and four columns of $\mathbf{B}$.
Through the \gls{ssr} extension, elements of $\mathbf{A}$ and $\mathbf{B}$ are prefetched from memory and streamed directly into registers \lstinline{ft0} and \lstinline{ft1}, respectively.
As a side effect, the two loops over the M and N dimensions of the $\mathbf{C}$ matrix could be collapsed into a single outer loop of M$\times$N iterations.
The dot product loop over the K dimension is mapped to an FREP hardware loop, after peeling the first and last iterations.
Specifically, the first iteration employs \lstinline{fmul} instructions, to avoid resetting registers \lstinline{c0}-\lstinline{c3} to zero on every outer loop iteration, while the \lstinline{fmadd} instructions in the last iteration write back to \lstinline{ft2}, forwarding results directly to memory via an \gls{ssr}.
After applying these optimizations, most of the overhead instructions in the \gls{matmul} loop are eliminated, enabling near-ideal issue rates of one instruction per cycle to the \gls{fpu}, with an energy-efficient single-issue core.

At a higher level in the memory hierarchy, double buffering is employed to overlap data movement and computation, with the \gls{dma} engine loading the next blocks of $\mathbf{A}$ and $\mathbf{B}$ to the TCDM, and storing the previously computed block of $\mathbf{C}$ to main memory, while the compute cores process the current blocks.
This ensures that the \glspl{fpu} are never idle waiting for $\mathbf{A}$, $\mathbf{B}$ and $\mathbf{C}$ to be moved between TCDM and main memory.

\section{Implementation}

In this section, we introduce and address the remaining inefficiencies in \gls{matmul} execution on highly-optimized energy-efficient general-purpose clusters implementing stream registers and zero-overhead loops.
This effort unfolds into two contributions, detailed in the following subsections.

\subsection{Zero-Overhead Nested Loops}

While the FREP mechanism enables zero-overhead processing of the innermost \gls{matmul} loop, iterating over the K dimension for dot product accumulation, the outer loop incurs two loop management instructions per iteration, resulting in an asymptotic $\frac{2}{K*\text{unroll}}$ cycle overhead (possibly more on pipelined processors).
For large \gls{matmul} sizes found in typical \glspl{llm}, featuring large K, the issue could be overlooked as the overhead seemingly becomes negligible.
However, the actual problem size executed within a cluster is only a tile of the full-size problem, as limited by L1 cache capacity.
Assuming a square problem size to optimize arithmetic intensity, problem sizes of 32$\times$32$\times$32 are common.

To address this inefficiency, we generalize the FREP extension to support loop nests matching the following template, including both perfectly and imperfectly nested loops:
\begin{lstlisting}[language=Python]
for i in range(N):      # frep N, A+B+C
    [..]                # A insns, with A >= 0
    for j in range(M):  # frep M, B
        [..]            # B insns, with B >= 1
    [..]                # C insns, with C >= 0
\end{lstlisting}
The outer \gls{matmul} loop can then also be mapped to an FREP instruction, eliminating all loop overheads in a \gls{matmul} tile computation.
Additionally, instructions in the outer loop can be fetched from the FREP buffer, instead of the instruction cache, reducing overall energy consumption.
We present the results of this optimization in \Cref{sec:results}, and compare to existing zero-overhead nested loop implementations in \Cref{sec:related-work-zonl}.
In this section, we present our generalized FREP implementation.

\begin{figure}[t!]
    \centering
    \includegraphics[width=\columnwidth]{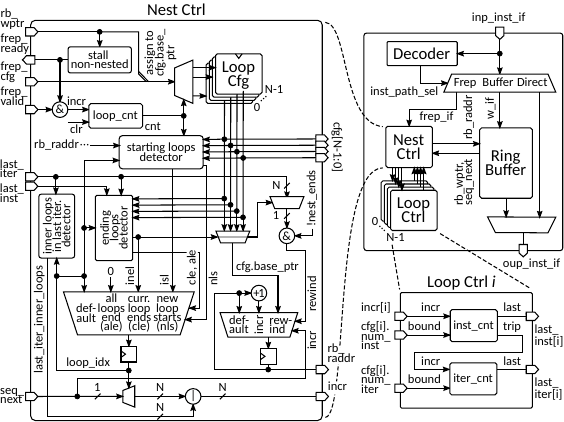}
    \caption{Block diagram of the FREP sequencer.}
    \vspace{-4pt}
    \label{fig:sequencer}
    \vspace{-8pt}
\end{figure}

\Cref{fig:sequencer} shows a block diagram of the FREP sequencer module, that receives instructions from the control core (\lstinline{inp_inst_if}) and forwards them to the \gls{fpu} (\lstinline{oup_inst_if}).
Instructions are partially decoded and binned into categories which undergo separate handling: 1) FREPs, 2) instructions which can be part of an FREP loop body and 3) instructions featuring a source or destination register in the integer \gls{rf}.
The latter bypass the sequencer logic and are forwarded directly to the \gls{fpu}, while instructions in the second category are stored in a \gls{rb}, from which they can be re-issued if they are part of an FREP body.
The \gls{rb} issues one instruction per cycle, whenever it is not empty (\lstinline{seq_next}).

On the other hand, FREP instructions are fully decoded to extract the number of instructions in the loop body and number of iterations\footnote{We retain the original instruction encoding proposed in \cite{zaruba2021}.}.
This information (\lstinline{frep_cfg}) is forwarded to the \textit{nest controller}, where it is stored together with the \gls{rb}'s current write pointer (\lstinline{rb_wptr}), marking the start of the loop. The read pointer (\lstinline{rb_raddr}) will be reset to this location when the loop ends.
Additionally, a loop counter (\lstinline{loop_cnt}) is incremented, to keep track of the depth of the loop nest which is dynamically constructed from incoming FREP instructions.

The maximum depth of a loop nest can be configured at design time by the \lstinline{N} parameter.
\lstinline{N} \textit{loop controllers} are instantiated to keep track of executed instructions (\lstinline{inst_cnt}) and iterations (\lstinline{iter_cnt}) for every loop and notify the nest controller when reaching the last instruction (\lstinline{last_inst}) and/or last iteration (\lstinline{last_iter}).
The nest controller aggregates this information to control which loops need to be incremented (\lstinline{incr}) on instruction issue (\lstinline{seq_next}).
To this end, it maintains an index of the currently active loop (\lstinline{loop_idx}), defined as the innermost loop containing the current instruction.
A loop's instruction counter ought to be incremented if and only if the loop is active (\lstinline{loop_idx==i}), or all inner loops are in their last iteration (\lstinline{last_iter_inner_loops[i]}), ensuring that instructions within inner loops are only counted once.

The active loop index is incremented when an inner loop is entered (\lstinline{nls}), and decremented when it is exited (\lstinline{cle}).
The first condition is evaluated by comparing the read pointer (\lstinline{rb_raddr}) and the next loop's base pointer (\lstinline{cfg[loop_idx+1].base_ptr}), the latter by checking if the loop is at the last instruction in its last iteration (\lstinline{last_iter&last_inst}).
As multiple loops may start and/or end on the same instruction, this logic has to be extended to detect the number of loops starting or ending on an instruction.
The \textit{starting loops detector} and \textit{ending loops detector} blocks, implemented around a leading and trailing zero counter, respectively, are dedicated to this task.
Starting from the currently active loop, the two modules respectively detect the innermost loop starting on the next instruction and the outermost loop ending on the next instruction, \textit{in a single cycle}.
This is a key contribution which allows us to sustain issue rates of one instruction per cycle on both perfectly and imperfectly nested loops, and sets our work apart from previous implementations, as discussed in \Cref{sec:related-work-zonl}. 
The loop index is then updated respectively to the innermost starting loop (\lstinline{isl}) or the innermost non-ending loop (\lstinline{inel}, i.e. the outermost ending loop - 1).
Finally, when an instruction is issued from the ring buffer, the read pointer (\lstinline{rb_raddr}) is incremented.
Only when the innermost non-ending loop is in its last instruction (\lstinline{last_iter[inel]}), and we are not in the last iteration of the outermost loop (\lstinline{!nest_ends}), the read pointer is reset (\lstinline{rewind}) to the base of the loop.

\subsection{Zero-Conflict Memory Subsystem}
\label{sec:interconnect}

As shown in \Cref{fig:gemm}, every compute instruction in the \gls{matmul} loop consumes two operands from memory.
The \lstinline{fmadd} instructions following the inner dot product loop, which occur every K cycles, additionally writeback their result to memory.
To sustain a throughput of one instruction per cycle, the memory must thus be able to serve two reads and one write, in the worst case, per cycle.
With 8 compute cores per cluster working on independent parts of the problem, this adds up to 16 reads and 8 writes, generically 24 requests, per cycle.

\begin{figure}[t!]
    \centering
    \includegraphics[width=\columnwidth]{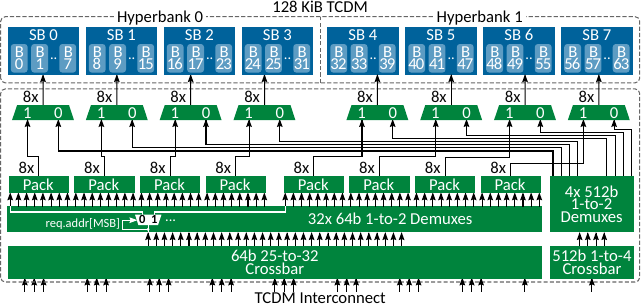}
    \vspace{-15pt}
    \caption{Block diagram of the zero-conflict memory subsystem and double buffering aware (Dobu) interconnect.}
    \label{fig:interconnect}
    \vspace{-8pt}
\end{figure}

As mentioned in \Cref{sec:background}, while every TCDM bank is single ported, up to 32 parallel 64-bit reads and writes are possible so long as simultaneous requests address different banks.
Thus, the bandwidth required by the cores can be attained by distributing the requests across $\ge$\,24 banks, achievable by laying out the $\mathbf{A}$, $\mathbf{B}$ and $\mathbf{C}$ matrices across 24 banks.

As further mentioned in \Cref{sec:background}, double buffering is employed to overlap computation and communication.
While the compute cores access the current blocks of $\mathbf{A}$, $\mathbf{B}$ and $\mathbf{C}$ in TCDM, the \gls{dma} engine will simultaneously be writing the next blocks of $\mathbf{A}$ and $\mathbf{B}$ into the TCDM and reading out the previously computed block of $\mathbf{C}$.
The buffers prepared (resp. read out) by the \gls{dma} must align with the layout assumed by the cores in the next (resp. previous) block iteration.
This means that if the core requests are spread out across 24 banks, the \gls{dma} requests will also be spread out across 24 banks, making it extremely difficult, if not impossible, to coordinate the 512-bit request of the \gls{dma} (spanning 8 banks) and the 24 64-bit requests of the cores (spanning 24 banks) to fully avoid bank conflicts across the available 32 banks.

Our solution to this problem involves instantiating 48 banks, divided among two \textit{hyperbanks} of 24 banks each.
The buffers prepared by the \gls{dma} in one hyperbank can potentially be accessed without conflicts by the cores, as the \gls{dma} prepares the buffers for the next iteration in the other hyperbank.
While this enables the 8 cores and the \gls{dma} to operate in parallel without conflicts, it results in a significant increase in area and routing congestion, as the all-to-all TCDM interconnect design scales poorly with the number of banks.

To solve this issue, we design a novel double-buffering-aware interconnect, referred to in the following as \textit{Dobu} interconnect, which can be adopted by general-purpose and specialized accelerators alike.
\Cref{fig:interconnect} shows a block diagram of the Dobu interconnect.
Like the original interconnect, it relies on a fully-connected crossbar to route requests from the cores to the banks within a hyperbank.
However, doubling the number of banks is achieved by doubling the number of hyperbanks (from 1 to 2), rather than increasing the number of banks per hyperbank.
A stage of demuxes is added after the fully-connected crossbar, to route requests to the correct hyperbank.
We choose an address mapping scheme which favours locality within each hyperbank: the TCDM is split into a contiguous address region per hyperbank, with interleaved addresses across banks in the hyperbank.
Each hyperbank is thus addressed akin to the original TCDM, with the address stripped of its most significant bit for hyperbank selection.

As we show in \Cref{sec:at-analysis}, this interconnect design allows us to scale up to \textit{64} banks, with little overhead.
In turn, this allows to serve even the most memory-intensive kernels as theoretically allowed by the RISC-V ISA\footnote{Featuring instructions with at most 3 source and 1 destination register, from which we derive $(\SI{3}{\reads} + \SI{1}{\write}) * \SI{8}{\cores} * 2 = \SI{64}{\banks}$.} on 8 cores, with the peak bandwidth required and without compromising performance when applying double buffering.

\begin{figure}[t!]
    \centering
    \includegraphics[width=0.90\columnwidth]{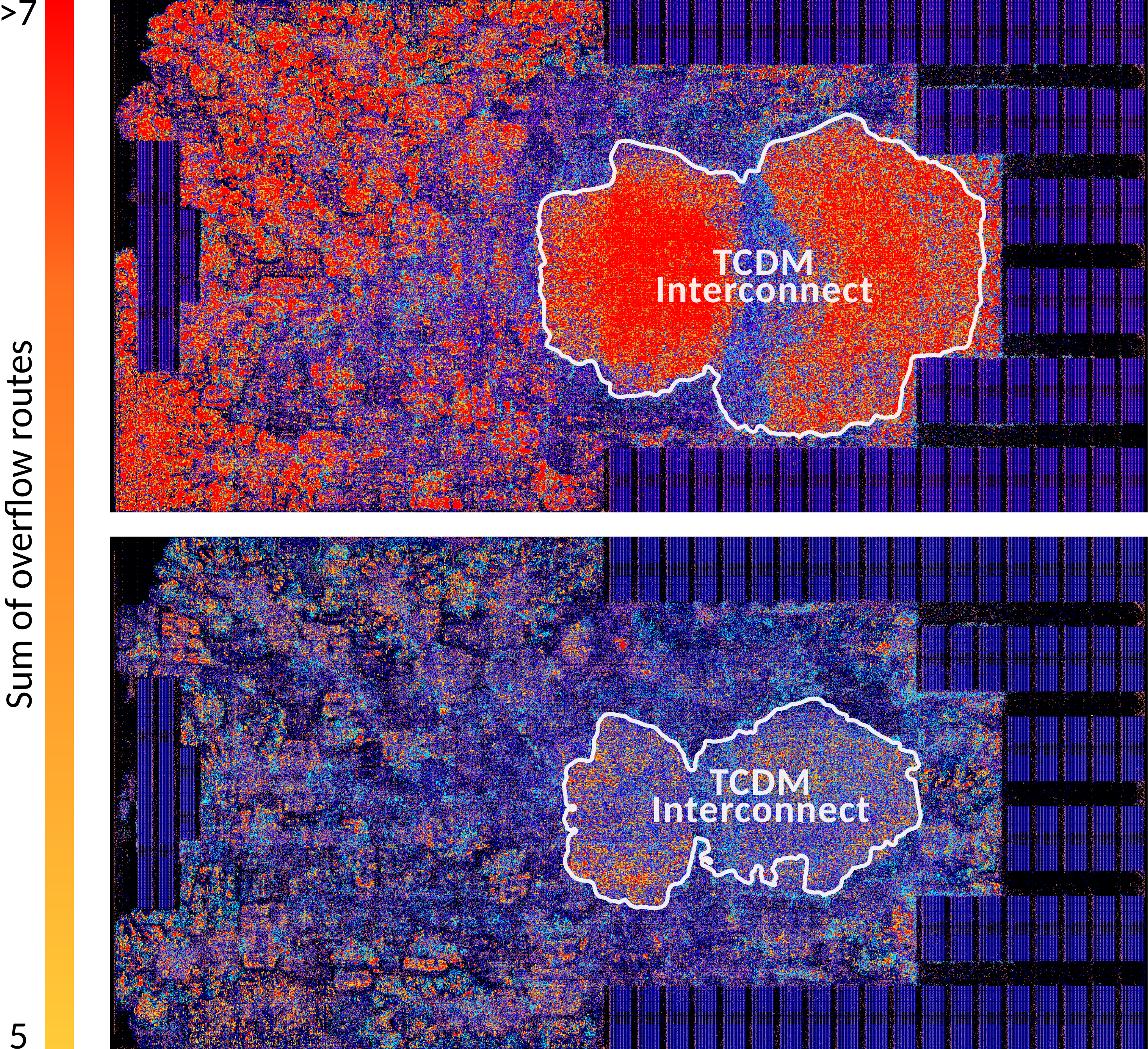}
    \vspace{-1pt}
    \caption{Placed-and-routed Snitch cluster in the \texttt{Zonl64fc} (top) and \texttt{Zonl64db} (bottom) configurations. Routing congestion is highlighted using the sum of overflow routes metric.}
    \label{fig:congestion}
    \vspace{-5pt}
\end{figure}

\section{Results}
\label{sec:results}

We implement the Snitch cluster in GlobalFoundries' 12LP+ FinFET technology using Fusion Compiler 2023.12, under a 1\,GHz target clock frequency constraint and worst-case conditions (SS, 125\,°C, 0.72\,V).
All area results are reported in \glspl{ge}, assuming $\SI{1}{\gegf} = \SI{0.121}{\micro\meter\squared}$.

Performance experiments are conducted at the target frequency in cycle-accurate RTL simulation using QuestaSim 2023.4.
Switching activities are extracted from post-layout simulations, and used for power estimation in PrimeTime 2022.03, under nominal conditions (TT, 25\,°C, 0.8\,V).

\subsection{Area and Timing Analysis}
\label{sec:at-analysis}

Implementation results for five variants of the Snitch cluster are presented in \Cref{tab:area}.
We assume the reference Snitch cluster from \cite{scheffler2025}, with a 128\,KiB 32-bank TCDM and fully-connected TCDM interconnect, as our baseline (\texttt{Base32fc}).
With exception of the \texttt{Zonl64fc} configuration, all variants meet the target frequency, as the critical paths lie within Snitch's \gls{fpu} and I\$.
The zero-overhead loop nest implementation (\texttt{Zonl32fc}) introduces a negligible \textless 3\% overhead on the overall cluster area.
To evaluate the interconnect-related cost of doubling the number of banks to 64, we fix the 128\,KiB memory size, halving the capacity of each memory macro.
The 64-bank fully-connected interconnect (\texttt{Zonl64fc}) takes a significant toll on cell area (+\ResultZonlSixtyFourFcCellAreaIncreaseOverZonlThirtyTwoFc\%) and wire length (+\ResultZonlSixtyFourFcWireLengthIncreaseOverZonlThirtyTwoFc\%), resulting in a heavily routing-congested design, as shown in \Cref{fig:congestion}.
The previous numbers exclude the additional area cost (+\ResultZonlSixtyFourFcMacroAreaIncreaseOverZonlThirtyTwoFc\%) associated to the smaller (less area-efficient) memory macros.
In comparison, our novel Dobu interconnect design significantly reduces interconnect complexity, resulting in a mere \ResultZonlSixtyFourDobuCellAreaIncreaseOverZonlThirtyTwoFc\% and \ResultZonlSixtyFourDobuWireLengthIncreaseOverZonlThirtyTwoFc\% increase in cell area and wire length, respectively, over \texttt{Zonl32fc}.
As shown in \Cref{fig:congestion}, our design is able to solve the severe routing congestion issues of the fully-connected interconnect.
Finally, we evaluate an extreme \gls{matmul}-optimized configuration featuring a 96\,KiB 48-bank TCDM\footnote{Despite the lower capacity, L1 data reuse is unaffected, as the data layout minimizing bank conflicts constrains every matrix within 8 banks \cite{yi2025}, featuring a constant 32\,KiB capacity.} with Dobu interconnect (\texttt{Zonl48db}).
Despite the 1.5$\times$ larger number of banks, the overall cell area (+\ResultZonlFourtyEightDobuCellAreaIncreaseOverZonlThirtyTwoFc\%) and wire length (-\ResultZonlFourtyEightDobuWireLengthIncreaseOverZonlThirtyTwoFc\%) are comparable to the baseline 32-bank memory subsystem configuration (\texttt{Zonl32fc}).
This is due to the reduced hyperbank width (from 32 to 24 banks), which results in smaller crossbars, the most area and routing-hungry components within the TCDM interconnect.
When accounting also for the \ResultZonlFourtyEightDobuMacroAreaIncreaseOverZonlThirtyTwoFc\% macro area reduction over \texttt{Zonl32fc}, the overall cluster area falls \ResultZonlFourtyEightDobuTotalAreaIncreaseOverZonlThirtyTwoFc\% below that of \texttt{Zonl32fc}.
We refer to our 48-bank configuration in the \gls{soa} comparison presented in \Cref{sec:soa}.

\begin{table}[t!]
    \begin{tabular*}{\columnwidth}{@{\extracolsep{\fill}} l c c c@{\hspace{2pt}} c c@{\hspace{2pt}} c @{}}
        \toprule
        Configuration & Cell area  & Macro area  & \multicolumn{2}{c}{Wire length}  & \multicolumn{2}{c}{Total area} \\
        \midrule
        \texttt{Base32fc}            & \ResultBaseThirtyTwoCellArea       & \ResultBaseThirtyTwoMacroArea       & \ResultBaseThirtyTwoWireLength       &                                                                    & \ResultBaseThirtyTwoTotalArea       &                                                                      \\
        \texttt{Zonl32fc}            & \ResultZonlThirtyTwoFcCellArea     & \ResultZonlThirtyTwoFcMacroArea     & \ResultZonlThirtyTwoFcWireLength     & (+\ResultZonlThirtyTwoFcWireLenOverBaseThirtyTwoFc\%)              & \ResultZonlThirtyTwoFcTotalArea     & (+\ResultZonlThirtyTwoFcTotalAreaOverBaseThirtyTwoFc\%)              \\
        \texttt{Zonl64fc}            & \ResultZonlSixtyFourFcCellArea     & \ResultZonlSixtyFourFcMacroArea     & \ResultZonlSixtyFourFcWireLength     & (+\ResultZonlSixtyFourFcWireLenOverBaseThirtyTwoFc\%)              & \ResultZonlSixtyFourFcTotalArea     & (+\ResultZonlSixtyFourFcTotalAreaOverBaseThirtyTwoFc\%)              \\
        \texttt{Zonl64dobu}          & \ResultZonlSixtyFourDobuCellArea   & \ResultZonlSixtyFourDobuMacroArea   & \ResultZonlSixtyFourDobuWireLength   & (+\ResultZonlSixtyFourDobuWireLenOverBaseThirtyTwoFc\%)            & \ResultZonlSixtyFourDobuTotalArea   & (+\ResultZonlSixtyFourDobuTotalAreaOverBaseThirtyTwoFc\%)            \\
        \textbf{\texttt{Zonl48dobu}} & \ResultZonlFourtyEightDobuCellArea & \ResultZonlFourtyEightDobuMacroArea & \ResultZonlFourtyEightDobuWireLength & (\textbf{\ResultZonlFourtyEightDobuWireLenOverBaseThirtyTwoFc\%})  & \ResultZonlFourtyEightDobuTotalArea & (\textbf{+\ResultZonlFourtyEightDobuTotalAreaOverBaseThirtyTwoFc\%}) \\
        \bottomrule
    \end{tabular*}
    
    \caption{Area [MGE] and routing [mm] cost of different Snitch cluster configurations: \texttt{\{Zonl|Base\}} with or without zero-overhead loop nest support, \texttt{\{32|48|64\}} number of banks and \texttt{\{fc|db\}} with fully-connected or Dobu interconnect. Increments are relative to the \texttt{Base32fc} configuration.}
    \label{tab:area}
    \vspace{-5pt}
\end{table}

\subsection{Performance and Energy Evaluation}

\begin{figure}
    \begin{subfigure}{0.47\columnwidth}
        \centering
        \includegraphics[width=\textwidth]{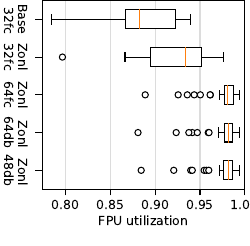}
    \end{subfigure}
    \hfill
    \begin{subfigure}{0.25\columnwidth}
        \centering
        \includegraphics[width=\textwidth]{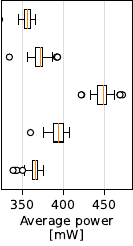}
    \end{subfigure}
    \hfill
    \begin{subfigure}{0.25\columnwidth}
        \centering
        \includegraphics[width=\textwidth]{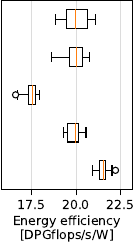}
    \end{subfigure}
    \vspace{-5pt}
    \caption{Box plots of the utilization attained on 50 randomly selected \gls{matmul} problem sizes and the respective average power consumption and energy efficiency.}
    \label{fig:results}
    \vspace{-9pt}
\end{figure}

To assess the impact of our optimizations on \gls{matmul} performance, following the methodology from \cite{yi2025}, we measure the \gls{fpu} utilization across 50 different problem sizes, randomly sampling $M, N, K \in \{8, 16, 24, \ldots, 120, 128\}$ with uniform distribution.
\Cref{fig:results} presents the results of this analysis, showing the distribution of the metrics of interest across all problem sizes, for different cluster configurations.
The utilization in the baseline (\texttt{Base32fc}) implementation ranges from \ResultBaseThirtyTwoFcMinUtilization\% to \ResultBaseThirtyTwoFcMaxUtilization\%, with a median of \ResultBaseThirtyTwoFcMedianUtilization\%.
The zero-overhead loop nest extension (\texttt{Zonl32fc}) improves this across all problem sizes, increasing the median up to \ResultZonlThirtyTwoFcMedianUtilization\%, with only a \ResultZonlThirtyTwoFcMedianPowerOverBaseThirtyTwoFc\% increase in power.
As the median energy consumption stays approximately constant, nested loops represent an effective low-power solution to increase performance.

\begin{table*}[t!]
    \begin{threeparttable}
        \setlength{\tabcolsep}{3pt}
        \begin{tabular*}{\textwidth}{@{\extracolsep{\fill}} l c c c c c c c c c c c c @{}}
            \toprule
            \multirow{3}{*}{\makecell{}}                 & \multicolumn{4}{c}{Area}                                                                                                                                                                                 & \multicolumn{4}{c}{Power}                                                                                                                                                                        & \multirow{3}{*}{\makecell{Util.}}                & \multirow{3}{*}{\makecell{Perf.}}     & \multirow{3}{*}{\makecell{Area\\Eff.}}   & \multirow{3}{*}{\makecell{Energy\\Eff.}}   \\
            \cmidrule(lr){2-5}
            \cmidrule(lr){6-9}
                                                         & Comp.                                 & L1 Mem.\,+\,Interco.                                                               & Ctrl.                                 & Total                               & Comp.                              & L1 Mem.\,+\,Interco.                                                        & Ctrl.                                  & Total                                &                                                  &                                       &                                          &                                            \\
            \midrule                                                                                                                                                                                                                                                                                                                                                                                                                                                                                                                                                                                                                                                        
            Ours [\texttt{Zonl48dobu}]                   & \ResultZonlFourtyEightDobuFpuArea     & \ResultZonlFourtyEightDobuTcdmArea\,+\,\ResultZonlFourtyEightDobuInterconnectArea  & \ResultZonlFourtyEightDobuControlArea & \ResultZonlFourtyEightDobuTotalArea & \ResultZonlFourtyEightDobuFpuPower & \ResultZonlFourtyEightDobuTcdmPower\,+\,\ResultZonlFourtyEightDobuTcdmPower & \ResultZonlFourtyEightDobuControlPower & \ResultZonlFourtyEightDobuTotalPower & \ResultZonlFourtyEightDobuUtilization\%          & \ResultZonlFourtyEightDobuPerformance & \ResultZonlFourtyEightDobuAreaEfficiency & \ResultZonlFourtyEightDobuEnergyEfficiency \\
            Snitch [\texttt{Base32fc}]                   & \ResultBaseThirtyTwoFcFpuArea         & \ResultBaseThirtyTwoFcTcdmArea\,+\,\ResultBaseThirtyTwoFcInterconnectArea          & \ResultBaseThirtyTwoFcControlArea     & \ResultBaseThirtyTwoFcTotalArea     & \ResultBaseThirtyTwoFcFpuPower     & \ResultBaseThirtyTwoFcTcdmPower\,+\,\ResultZonlFourtyEightDobuTcdmPower     & \ResultBaseThirtyTwoFcControlPower     & \ResultBaseThirtyTwoFcTotalPower     & \ResultBaseThirtyTwoFcUtilization\%              & \ResultBaseThirtyTwoFcPerformance     & \ResultBaseThirtyTwoFcAreaEfficiency     & \ResultBaseThirtyTwoFcEnergyEfficiency     \\
            OpenGeMM\cite{yi2025}\textsuperscript{\ddag} & \ResultOpenGemmComputeArea*           & \ResultOpenGemmTcdmArea                                                            & \ResultOpenGemmControlArea            & \ResultOpenGemmTotalArea            & \ResultOpenGemmFpuPower*           & \ResultOpenGemmTcdmPower                                                    & \ResultOpenGemmControlPower            & \ResultOpenGemmTotalPower            & \ResultOpenGemmUtilization\%\textsuperscript{\S} & \ResultOpenGemmPerformance            & \ResultOpenGemmAreaEfficiency            & \ResultOpenGemmEnergyEfficiency            \\
            \bottomrule
            \addlinespace[2pt]
        \end{tabular*}
        \begin{tablenotes}
            \footnotesize
            \item[\ddag]Power is scaled by 4.92$\times$ to account for technology (0.7$\times$), voltage and frequency differences ($\propto V^2f$); area assumes $\SI{1}{\getsmc} = \SI{0.138}{\micro\meter\squared}$.
            \item[\S]While power results are referred to a 32$\times$32$\times$32 \gls{matmul}, the authors only report the median utilization over a range of randomly sized matrices.
            \item[*]Derived from \texttt{Base32fc}, subtracting the \gls{dma} core's \gls{fpu} engine, and scaling by the \gls{fpu} utilization.
        \end{tablenotes}
    \end{threeparttable}
    \vspace{-3pt}
    \caption{Comparison of the area [MGE] and power [mW] breakdowns, performance [DPGflop/s], area efficiency [DPGflop/s/mm\textsuperscript{2}] and energy efficiency [DPGflop/s/W] of our optimized cluster, the baseline Snitch cluster and OpenGeMM.}
    \label{tab:soa}
    \vspace{-5pt}
\end{table*}

By doubling the number of banks to 64 (\texttt{Zonl64fc}), we can eliminate all TCDM conflicts, as described in \Cref{sec:interconnect}.
As a result, the median \gls{fpu} utilization increases by an additional \ResultZonlSixtyFourFcMedianUtilizationOverZonlThirtyTwoFc\%, up to \ResultZonlSixtyFourFcMedianUtilization\%.
Excluding a few outliers, none of which fall below \ResultZonlSixtyFourFcMinUtilization\%, we attain between \ResultZonlSixtyFourFcBottomWhiskerUtilization\% and \ResultZonlSixtyFourFcTopWhiskerUtilization\% \gls{fpu} utilization on all problem sizes.
However, this solution introduces a \ResultZonlSixtyFourFcMedianEnergyOverZonlThirtyTwoFc\% increase in median energy consumption, owing to the increased interconnect complexity.
We developed the Dobu interconnect to address this issue.

In terms of performance, our 64-bank double buffering aware interconnect (\texttt{Zonl64db}) attains comparable utilizations to the fully-connected implementation (\texttt{Zonl64fc}).
At the same time, we can reduce the energy overhead to zero, as the small cost from the increased interconnect complexity is balanced out by the savings on wasteful TCDM conflicts.
The \ResultZonlSixtyFourDobuMedianPowerOverZonlThirtyTwoFc\% overhead in power can be attributed to the \ResultZonlSixtyFourDobuMedianUtilizationOverZonlThirtyTwoFc\% increase in utilization over \texttt{Zonl32fc}.
With the Dobu interconnect, we can thus harvest the performance benefits from eliminating bank conflicts, without compromising energy efficiency, proving it an effective low-power technique to eliminate memory inefficiencies on double-buffered workloads.

Finally, the 48-bank implementation with Dobu interconnect (\texttt{Zonl48db}) attains similar utilizations to its 64-bank counterparts, whilst improving the median energy efficiency by \ResultZonlFourtyEightDobuMedianEnergyOverBaseThirtyTwoFc\%, compared to the baseline cluster (\texttt{Base32fc}).

\section{Related Work}
\label{sec:related-work}



\subsection{Zero-overhead loops}
\label{sec:related-work-zonl}

Early works introduced zero-overhead loops in the context of \glspl{dsp} already in the late 1990s \cite{bajwa1997, uh1999}.
Successive studies extended this line of work to support nested loops, but either exclusively target perfectly nested loops \cite{talla2003, kavvadias2010} or trade off performance to also support imperfectly nested loops \cite{tsao2003, gautschi2017, kavvadias2008, wei2010, vadivel2017}.
The primary difficulty lies in handling nested loops which start and/or end at the same instruction, as this requires detecting how many loops start or end on a given instruction, to update the active loop index.
Works targeting perfectly nested loops can rely on the assumption that all loops in the nest share the same loop body, i.e. all start and end on the same instructions, simplifying the design.
Conversely, no work supporting imperfectly nested loops handles the detection of loops starting or ending on the same instruction, \textit{in a single cycle}.
Instead, they either 1) do not support loops which start and/or end on the same instruction \cite{gautschi2017}, 2) iteratively detect such loops over multiple cycles negatively affecting performance \cite{kavvadias2008, vadivel2017}, or 3) do not provide a detailed account if and how such cases are handled \cite{tsao2003, wei2010}.
In contrast to previous efforts, our work presents an approach to support both perfectly and imperfectly nested loops with zero-cycle overhead, and demonstrates its importance in achieving optimal \gls{matmul} performance.


\subsection{Shared memory cluster interconnect}

Memory banking is a well known technique to increase memory bandwidth.
Several works adopt banked scratchpad memories as a communication means and shared L1 data cache (TCDM) between cores in energy-efficient clusters \cite{gautschi2017, zaruba2021}, interconnecting memory banks and cores through all-to-all networks \cite{rahimi2011}.
Prior works have also developed heterogeneous TCDM interconnects (HCI) \cite{prasad2023} based on a similar core-to-memory interconnect as adopted by Snitch, integrating an additional shallow interconnect branch to provide low-latency, high-bandwidth access to specialized accelerators (akin to the \gls{dma} XBAR in Snitch).
Their approach introduces multiplexing logic to arbitrate TCDM accesses between interconnect branches, in the same way \gls{dma} accesses are arbitrated in the Snitch cluster \cite{scheffler2025}, and is thus subject to memory conflicts between the core and accelerator accesses.
In contrast, we extend the interconnect to scale with the number of TCDM banks, thus enabling a \textit{zero-conflict} memory subsystem, by adding demultiplexing logic between each interconnect branch and multiple hyperbanks.
Finally, as noted by Gautschi et al. \cite{gautschi2017}, increasing the number of banks can negatively impact power consumption, as the power per TCDM access grows with interconnect complexity.
With our Dobu interconnect design, we demonstrate a solution to increase the number of banks, without significantly affecting power consumption.
To the best of our knowledge, our work is the first to demonstrate a software-aware solution to this problem, by specializing the interconnect for a specific (yet widespread) application pattern (double buffering), which allows to reduce TCDM contention on double-buffered applications in an energy-efficient manner.
While we demonstrate our work on a cluster of scalar cores, our solution can potentially benefit any architecture employing double buffering, including clusters of vector and matrix units.


\subsection{Energy-efficient shared memory clusters}
\label{sec:soa}

Most \gls{ml} accelerators feature a common architectural template, consisting of a tile-based fabric of compute clusters equipped with local L1 memory \cite{prabhakar2024, zhang2024, xu2024, paik2024, maddury2024, vasiljevic2024}.
Diverse compute cluster architectures have been explored in the literature: from streamlined general-purpose multicore clusters \cite{zaruba2021}, where every \gls{pe} can be individually programmed, to compact vector engines \cite{perotti2025} and specialized matrix multiplication units \cite{yi2025}, which trade-off flexibility and programmability for efficiency.

\Cref{tab:soa} compares our optimized cluster (\texttt{Zonl48dobu}), the baseline Snitch cluster (\texttt{Base32fc}) and the \gls{soa} OpenGeMM \gls{matmul} accelerator \cite{yi2025}, on a 32$\times$32$\times$32 \gls{matmul} kernel.
We break down area and power into four contributions, related to different microarchitectural functions: compute, memory, interconnect and control.
For an arithmetic-precision-agnostic comparison, we replace the area and power of OpenGeMM's 8$\times$8$\times$8 INT8 GEMM core, with figures derived for a 2$\times$2$\times$2 FP64 SIMD-capable GEMM core.
All clusters feature an equal peak performance of 8\,DPGflop/s.

By eliminating control and memory-induced inefficiencies through our extensions, we are able to close the utilization and performance gap with the \gls{soa} specialized OpenGeMM accelerator, with only \ResultZonlFourtyEightDobuEnergyEfficiencyOverOpenGemm\% difference in energy efficiency, while offering individually programmable \glspl{pe} which can be employed on a significantly larger range of workloads.

\section{Conclusion}

In this work, we propose an energy-efficient RISC-V cluster architecture for \gls{ml} acceleration.
Building on the \gls{soa} Snitch cluster, we develop two general-purpose, lightweight extensions to eliminate control and memory access induced inefficiencies in \gls{matmul} workloads, without increasing processor complexity or sacrificing programmability.
Specifically, we propose a ``zero-overhead loop nest'' extension to eliminate outer loop handling instructions, and a ``zero-conflict memory subsystem'', leveraging a novel double-buffering-aware interconnect, to eliminate bank conflicts.
Through these enhancements, we attain near-ideal \gls{fpu} utilizations between \ResultZonlFourtyEightDobuBottomWhiskerUtilization\% and \ResultZonlFourtyEightDobuTopWhiskerUtilization\%, 
improving the performance and energy efficiency of the heavily-optimized baseline Snitch cluster by \ResultZonlFourtyEightDobuMedianUtilizationOverBaseThirtyTwoFc\% and \ResultZonlFourtyEightDobuMedianEnergyOverBaseThirtyTwoFc\%, respectively.
We achieve comparable utilization and performance to OpenGeMM \cite{yi2025}, a specialized \gls{soa} accelerator, with only \ResultZonlFourtyEightDobuEnergyEfficiencyOverOpenGemm\% difference in energy efficiency, while providing a fully-programmable general-purpose architecture which can support a significantly wider range of workloads.

\section*{Acknowledgment}

\ifdefined\blindreview
Hidden for double blind review.
\else
This work has been supported in part by ‘The European Pilot’ project under grant agreement No 101034126 that receives funding from EuroHPC-JU as part of the EU Horizon 2020 research and innovation programme.
\fi

\bibliography{paper}
\bibliographystyle{IEEEtran}

\end{document}